\documentclass[
superscriptaddress,
preprint,
 amsmath,amssymb,
 aps,
 prc
]{revtex4-1}

\usepackage{graphicx}
\usepackage{dcolumn}
\usepackage{bm}
\usepackage{soul,xcolor}
\usepackage{amssymb}
\usepackage{mathrsfs}
\usepackage[pdfencoding=auto,psdextra]{hyperref}
 \usepackage{CJK}

\setstcolor{blue}

\begin{document}
 \begin{CJK*}{UTF8}{}

\title{Pairing phase transition in the odd-A nuclei {:} identification and classification }

\author{Yumeng Wang (\CJKfamily{gbsn}王宇萌)}
\affiliation{School of Science, Jiangnan University, Wuxi 214122, China.}

 \author{Yuhang Gao (\CJKfamily{gbsn}高宇航)}
\affiliation{School of Science, Jiangnan University, Wuxi 214122, China.}

 \author{Lang Liu (\CJKfamily{gbsn}刘朗)}
\email{liulang@jiangnan.edu.cn}
\affiliation{School of Science, Jiangnan University, Wuxi 214122, China.}

\begin{abstract}

The investigation into the pairing phase transition in the odd-A nucleus $^{161}\mathrm{Dy}$ utilizes a sophisticated blend of covariant density functional theory and the shell-model-like approach. It is discerned that variations in thermodynamic quantities at the critical temperature do not exclusively align with pairing phase transitions. The presence of an S-shaped heat capacity curve, often interpreted as indication of such transitions, does not offer a definitive confirmation. Additional factors, including the blocking effect, can modify the heat capacity curve and impede the transition process. The pairing phase transition in $^{161}\mathrm{Dy}$, occurring around 0.7 to 1.0 MeV, is unequivocally characterized as a first-order transition. Furthermore, an exploration into the impact of varying strengths of pairing correlations on these transitions reveals a nonlinear relationship, adding complexity to the transition dynamics.

\end{abstract}

\maketitle
\end{CJK*}

\section{Introduction}
\label{sec:1}
The pairing phase transition is widely recognized as essential for comprehending the thermodynamic properties of atomic nuclei~\cite{Bohr1998a}. Due to the precise measurement of level densities at low excitation energies, thermodynamic quantities have been deduced within the canonical ensemble framework, thereby confirming the S-shaped curve of heat capacity as a function of temperature~\cite{PhysRevC.63.021306,PhysRevC.63.044309,PhysRevLett.83.3150,PhysRevC.68.034311}. The significance of this S-shaped curve has been extensively explored across various theoretical models, including the nuclear shell model~\cite{PhysRevC.58.3295,PhysRevLett.87.022501,LANGANKE2005360}, the mean field model~\cite{PhysRevLett.85.26,PhysRevC.62.044307,PhysRevC.61.044317,PhysRevC.88.034308,Li2015}, and others~\cite{PhysRevC.63.044301,Guttormsen2001}. It is considered the hallmark of the pairing phase transition from the superfluid phase to the normal phase. However, in the superfluid phase, particle number conservation is violated by the BCS theory or the Bogoliubov transformation within the mean field model. The importance of conservation laws in studying phase transitions of finite systems has been elucidated in Refs.~\cite{PhysRevC.72.044303,Gambacurta2013,Esebbag1993}. To address these drawbacks, we employ the shell-model-like approach, which ensures particle number conservation and accurately deals with the blocking effect~\cite{Zeng1983,Zeng1994,Meng2006}. Additionally, covariant density functional theory (CDFT) has effectively described numerous nuclear phenomena~\cite{Ring1996,Vretenar2005a,Meng2006a,Meng2013}. The shell-model-like approach has been integrated within the CDFT framework~\cite{Meng2006}. Previous studies have examined the thermodynamic properties of even-even nuclei and odd-A nuclei using CDFT and the shell-model-like approach~\cite{PhysRevC.92.044304,Yan_2021}.

Although the pairing and pairng phase transition has been extensively investigated in many aspects~\cite{Horoi2007,Gao2023a,Cai2023,Tajima2019}, few studies have focused on the order of these transitions, and it remains to be more clearly defined. The initial classification of general types of phase transitions, introduced by Paul Ehrenfest in 1933, has played a pivotal role in the thermodynamic study of critical phenomena~\cite{10.2307/41134053}. Subsequently, Lee and Yang proposed a theorem on the distribution of roots of the grand partition function, predicting its broad applicability even in small systems~\cite{PhysRev.87.410}. In Refs.~\cite{PhysRevLett.84.3511,PhysRevA.64.013611}, a classification scheme for phase transitions in finite systems was proposed, based on the distribution of zeros (DOZ) of the canonical partition function in complex temperature. This method has been successfully applied to determine the order of phase transitions in $^{162}\mathrm{Dy}$~\cite{Gao2023}. To our knowledge, it has not yet been utilized to study the order of the pairing phase transition in odd-A nuclei.

This work aims to further investigate the microscopic mechanism underlying the S-shaped heat capacity curve, examining whether its appearance entirely corresponds to the occurrence of pairing phase transitions. If a pairing phase transition does occur, the order of the transition in odd-A nuclei remains unclear, and the influence of the blocking effect on the phase transition needs to be studied. Additionally, this research will explore whether and how the strength of the pairing correlation affects the pairing phase transition, an aspect that has not been addressed in previous studies.

In this work, we use $^{161} \mathrm{Dy}$ as an example. By employing the shell-model-like approach and covariant density functional theory, combined with the effective interaction PC-PK1~\cite{Zhao2010}, we investigate the thermodynamic properties and the influence of the blocking effect on the pairing phase transition. The order of the pairing phase transition is determined according to a classification scheme based on the distribution of zeros of the pairing partition function in the complex temperature plane. Furthermore, we construct an ideal odd system and an ideal even system to explore the impact of pairing correlation strength on the pairing phase transition, as evidenced by changes in the order of the transition.

This paper is organized as follows: In $\text{Sec.II}$, we briefly introduce the theoretical framework. $\text{Sec.III}$ presents the results and discussion. Finally, $\text{Sec.IV}$ contains the summary.

\section{Theoretical framework}
\label{sec:2}

In this section, we introduce briefly the basic formalism of the CDFT theory, which has been summarized in Ref.\cite{Zhao2010}. CDFT begins with a Lagrangian. From this Lagrangian, the Dirac equation for nucleons with local scalar $S(\bm{r})$ and vector $V^{\mu}(\bm{r})$ potentials can be deduced as 
\begin{equation}
\left[ \gamma_{\mu} (i \partial^{\mu} - V^{\mu}) - (m + S) \right] \psi_{\xi} = 0,
\label{eq:1}
\end{equation}
where
\begin{equation}
S(\bm{r}) = \Sigma_{S},  V(\bm{r}) = \Sigma^{\mu} + \vec{\tau} \cdot \vec{\Sigma}^{\mu}_{TV},
\label{eq:2}
\end{equation}
and $\psi_{\xi}$ is Dirac spinor.

In this study, SLAP is applied within the CDFT framework to deal with the pairing correlations by diagonalizing the Hamiltonian in the multi-particle configurations (MPCs) space, which is constructed by the single-particle levels obtained through the CDFT method. The total Hamiltonian reads
\begin{align}
  H &= H_{\rm s.p.} + H_{\rm pair} \notag
 \\
    &= \sum\limits_{i}\varepsilon_{i}a^{+}_{i}a_{i} -G\sum\limits^{i\neq j}_{i,j>0} a^{+}_{i}a^{+}_{\bar{i}}a_{\bar{j}}a_{j}, 
\label{eq:3} 
\end{align}
where $\varepsilon_{i}$ is the single-particle energy gained from the Dirac equation (\ref{eq:1}), $\bar i$ is the time-reversal state of $i$, and $G$ represents constant pairing strength.

For a system with an odd particle number $N = 2n+1$, it is necessary to block the state occupied by the odd nucleons. The MPCs for $s$ = 0 states can be expressed as,
\begin{align}
	|c_bc_1\bar{c}_1\cdots c_n\bar{c}_n\rangle=a_b^+(a_{c_1}^+a_{\bar{c}_1}^+\cdots a_{c_n}^+a_{\bar{c}_n}^+)|0\rangle,
	\label{eq:4}
\end{align}
in which $b$ denotes the single particle level blocked by the odd nucleon.

During the calculation process, a cutoff energy $E_c$ is needed to truncate the MPCs space. The configurations with energies $E_m-E_0 \leq E_c$ are used to diagonalize the Hamiltonian~(\ref{eq:3}), in which $E_m$ and $E_0$ are the energies of the $m$th configuration and the ground-state configuration, respectively. 

After the diagonalization of the Hamiltonian (\ref{eq:3}), the nuclear many-body wave function is shown as
\begin{align}
|\psi_\beta\rangle =&\sum\limits_{c_{1}\cdots c_{n}}{v_{\beta,\,c_1\cdots c_n}}|c_1\bar{c}_1\cdots c_n\bar{c}_n\rangle  \notag \\
& +\sum\limits_{i,j}{\sum\limits_{c_{1}\cdots c_{n-1}}{v_{\beta(ij),\,c_1\cdots c_{n-1}}}|i\bar{j}c_1\bar{c}_1\cdot\cdot\cdot c_{n-1}\bar{c}_{n-1}\rangle} \notag \\
& + \cdots,
\label{eq:5}
\end{align}
where $\beta = 0$ corresponds the ground state, $\beta = 1, 2, 3, \ldots$ correspond the excited states with the excitation energy $E_{\beta}$, and $v_{\beta}$ means the coefficient after diagonalization.

Moreover, the pairing gap can be calculated by Refs.~\cite{Canto1985,Egido1985,RevModPhys.61.131}
\begin{align}
\Delta_\beta=G\left[-\frac{1}{G}\langle\Psi_\beta|H_{\mathrm{p}}|\Psi_\beta\rangle\right]^{1/2},
	\label{eq:7}
\end{align}
The thermodynamic quantities of the pairing interaction are calculated in the canonical ensemble~\cite{PhysRevC.76.024319}. The partition function $Z$ can be calculated by
\begin{equation}
Z = \sum\limits^{\infty}_{\beta=0}\eta(E_{\beta})\,e^{-E_{\beta}/T}.
\label{eq:8}
\end{equation}
where $E_{\beta}$ is the excitation energy which could be acquired through the CDFT + SLAP method, and the corresponding level density $\eta (E_{\beta})$ is taken as $2^s$, i.e., the degeneracy of each state.

The classification scheme for phase transitions relies completely on Ref.~\cite{PhysRevC.66.024322}. Only the main formulas of the classification scheme will be given below. The scheme relies on the DOZ of the canonical partition function in the complex temperature plane.

The inverse complex temperature is defined as:
\begin{equation}
\mathcal{B} = \beta + i\tau,
\label{eq:9}
\end{equation}
where $\beta=1/T$ and $\tau$ corresponds the imaginary part of the inverse complex temperature which is measured in $\mathrm{MeV^{-1}}$. The zeros of the canonical partition function typically line up on curves in the complex temperature plane. The zeros are denoted by $(\beta_{j},\tau_{j})$. If there are $n$ zero points, $j$ = 1 $\ldots$ $n$ and $j$ increases with the increasing distance from the real axis.

If the first three zeros closest to the real axis have been determined, the average inverse distance between zeros can be calculated as
\begin{equation}
\Phi(\widetilde{\tau}_j)=\dfrac{1}{d_j},
\label{eq:10}
\end{equation}
where $\tilde{\tau}_{j}=(\tau_{j}+\tau_{j+1})/2$, $d_{j}=\sqrt{(\beta_{j+1}-\beta_{j})^{2}+(\tau_{j+1}-\tau_{j})^{2}}$. The function $\Phi$ can then be approximated in the vicinity of the real axis by a power law of $\tau_{j}$ only,
\begin{equation}
\Phi(\tau_j)\propto\tau_j^\alpha,
\label{eq:11}
\end{equation}
then $\alpha$ can be calculated by means of
\begin{equation}
\alpha=\dfrac{\ln\Phi(\tau_3)-\ln\Phi(\tau_2)}{\ln\tau_3-\ln\tau_2}.
\label{eq:12}
\end{equation}
According to the definition in Ref.~\cite{PhysRevC.66.024322}, 
when $\alpha$ $<$ 0, the system displays a first-order phase transition. When 0 $<$ $\alpha$ $<$ 1, the phase transition is of second order. When $\alpha$ $>$ 1, higher-order phase transition occurs.

\section{Results and discussion}
\label{sec:3}

\subsection{$^{161} \mathrm{Dy} $ }

For neutrons and protons, 16 single-particle levels and 9 valence nucleons are selected to construct a multi-particle configuration space, including 5 levels on and below the Fermi surface and 11 energy levels above the Fermi surface. PC-PK1 is chosen as the effective interaction. For neutrons, the Fermi surface is occupied by a single particle, and the maximum seniority number $s$ is 8. The span of the upper and lower energy levels of the model space is 7.56 $\mathrm{MeV}$. The corresponding pairing strengths of neutrons and protons are 0.29 MeV and 0.32 MeV, which can reproduce the experimental odd-even mass differences well. In this paper, the treatment of blocking is consistent with the method in Ref.~\cite{Yan_2021}. ``Bk1'' corresponds to the blocking of the Fermi surface, while ``bkn'' corresponds to the blocking of the (n-1)th energy level above the Fermi surface.


\begin{figure}[htbp]
\centering
 \includegraphics[scale=.20]{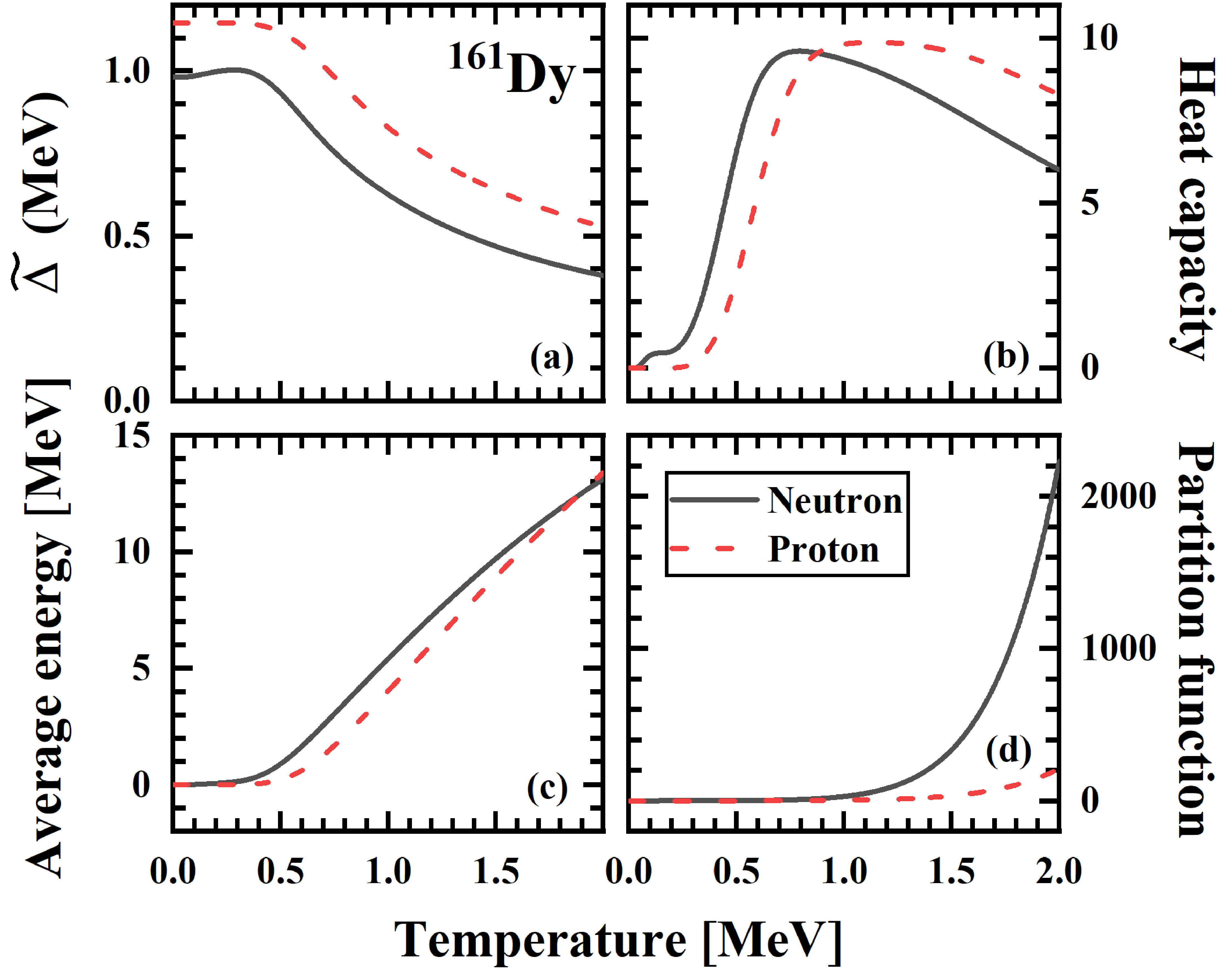}
\caption{(Color online) The pairing gap (a), heat capacity (b), average energy (c), and partition function (d) of neutrons (black solid lines) and protons (red dashed line) for $^{161} \mathrm{Dy} $ as functions of temperature.} 
\label{fig:example1}
\end{figure}

Figure~$\ref{fig:example1}$ illustrates the behavior of the pairing gap (a), heat capacity (b), average energy (c), and partition function (d) of neutrons (indicated by black solid lines) and protons (represented by red dashed lines) in $^{161} \mathrm{Dy}$ as functions of temperature. Subfigure (a) reveals obvious increase in the neutron curve at temperature below 0.3 MeV, contrasting with the almost constant trend observed in the proton curve within this temperature range. This behavior in the neutron curve indicates the pairing re-entrance phenomenon, wherein the pairing correlation in nuclei with an odd number of neutrons or protons is bolstered as temperature rises. This enhancement is primarily attributed to the influence of the single-particle level structure near the level occupied by the odd nucleons, which weakens the blocking effect and reinforces the pairing correlation.
Regarding heat capacity, both protons and neutrons exhibit an S-shaped curve. The initial turning point occurs at approximately 0.3 MeV for protons and 0.2 MeV for neutrons, aligning with the turning points of the pairing gap and average excitation energy. A second turning point emerges at around 1 MeV for protons and 0.7 MeV for neutrons, corresponding to the critical temperature of the phase transition. Additionally, the neutron curve demonstrates fluctuations in the low-temperature region. In the average energy diagram (c), both curves are nearly zero at low temperatures (around 0.25 MeV for neutrons and 0.3 MeV for protons), gradually increasing in a near-linear way thereafter.



\begin{figure}[ht!]
  \centering
    \includegraphics[scale=0.14]{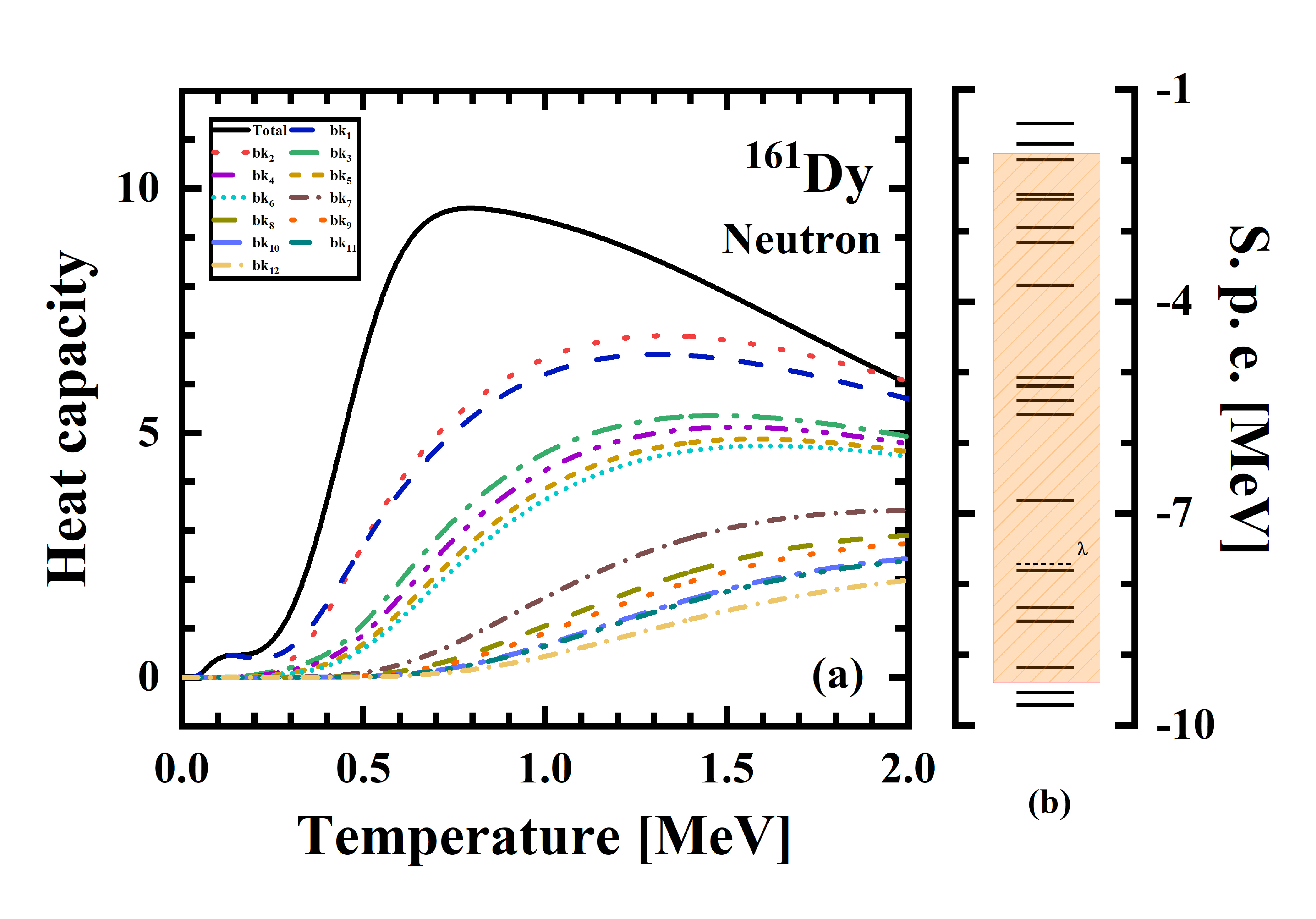}
\caption{(Color online) (a) The heat capacity of neutrons with the total and different levels blocked for $^{161} \mathrm{Dy} $ as functions of temperature. (b) The single particle levels of neutrons for $^{161} \mathrm{Dy} $. The levels in the orange area are used to construct the SLAP MPCs. $\lambda$ corresponds to the Fermi surface.} 
\label{fig:example2}
\end{figure}

Figure~$\ref{fig:example2}$ (a) displays the heat capacity of neutrons in $^{161} \mathrm{Dy}$, considering total and different blocking scenarios, as functions of temperature. The fluctuations observed at low temperatures primarily stem from the contribution of ``bk1", attributable to the single-particle level structure. In the case of ``bk2", corresponding to the blocking of the first level above the Fermi surface, Figure~$\ref{fig:example2}$ (b) illustrates that no levels exist around this energy level. Consequently, an energy input is required to excite nucleons, resulting in near-zero heat capacity at low temperatures. Subsequent blockings involve energy levels occupied by odd nucleons, which have less impact on the shape of the heat capacity curve. Hence, accurate treatment of Fermi surface blocking and its neighboring levels is crucial for investigating the relevant properties of odd-A nuclei.



\begin{figure}[ht!]
  \centering
    \includegraphics[scale=.13]{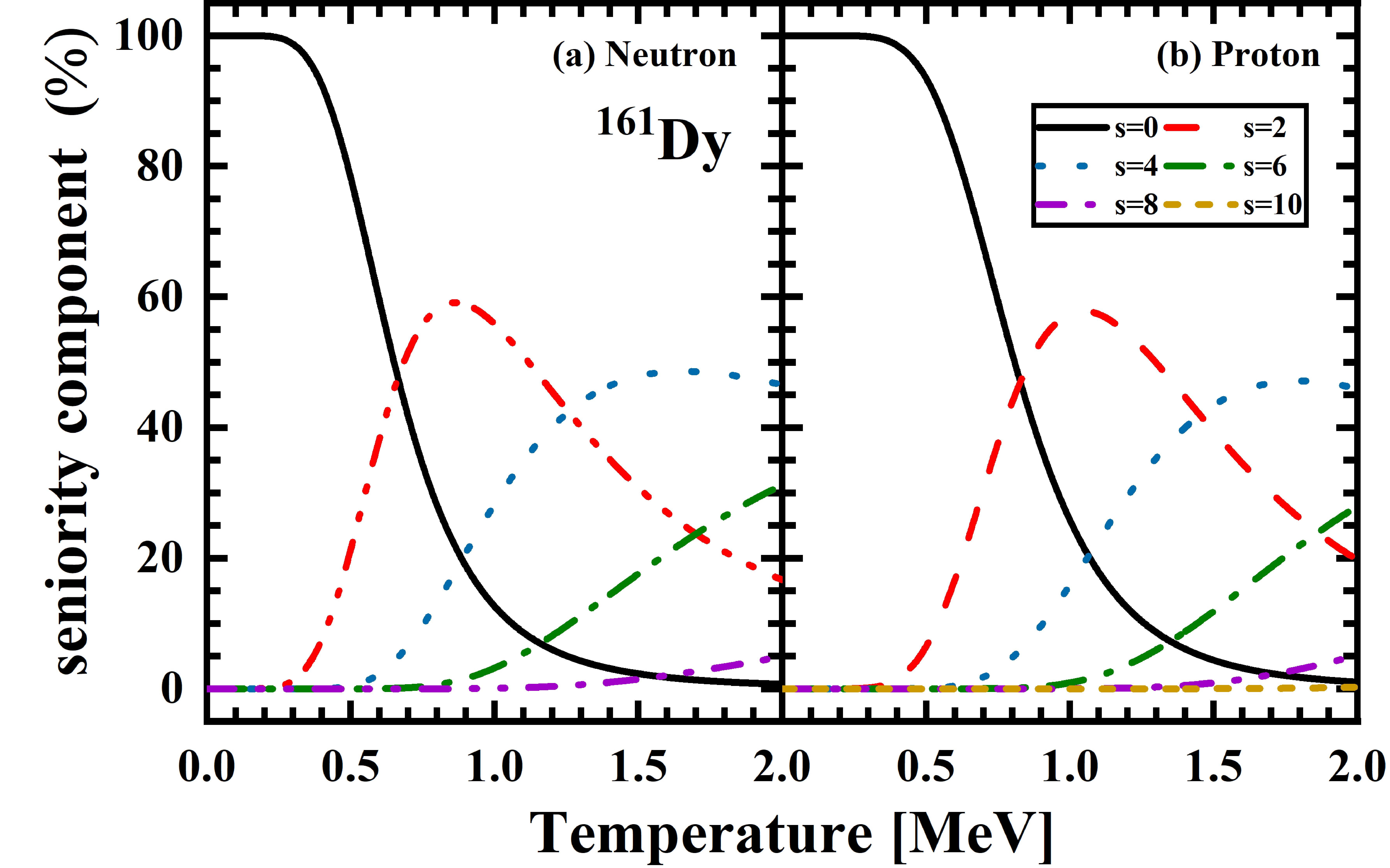}
\caption{(Color online) Seniority components with different seniority numbers $s$ = 0,2,4,6,8 of neutrons (a) and $s$ = 0,2,4,6,8,10 of protons (b) for $^{161} \mathrm{Dy} $ as functions of temperature.} 
\label{fig:example3}
\end{figure}

To gain a detailed microscopic understanding of the nuclear pairing transition, it is valuable to examine the number of Cooper pairs that become broken with increasing temperature in $^{161} \mathrm{Dy}$. In this analysis, we follow the definition of the seniority component as outlined in Ref.~\cite{PhysRevC.92.044304} 
\begin{align}
	\chi_s=Z^{-1}\sum_{\beta\in\{s\}}\eta(E_\beta)e^{-E_\beta/T}.
	\label{eq:7}
\end{align}
The seniority component illustrates the contribution of excited states with specific seniority numbers to the total average energy. Figure~$\ref{fig:example3}$ presents the seniority components for different seniority numbers $s$ = 0, 2, 4, 6, 8 of neutrons (a) and $s$ = 0, 2, 4, 6, 8, 10 of protons (b) in $^{161} \mathrm{Dy}$ as functions of temperature.
It is observed that $s$ = 0 states contribute nearly 100$\%$ below T $\approx$ 0.3 MeV for protons and 0.2 MeV for neutrons, aligning with the vanishing heat capacity illustrated in Fig.~\ref{fig:example1}. As the temperature exceeds 0.3 MeV for protons and 0.2 MeV for neutrons, the contribution of $s$ = 0 states begins to decrease. Simultaneously, the curve for $s$ = 2 states starts to rise, corresponding to the first turning point in the heat capacity curve and the temperature where the pairing gap curve begins to decline and the average energy curve begins to rise. The curves for $s$ = 2 peak at 0.8 MeV for neutrons and 1 MeV for protons, respectively, corresponding to the second turning point of the heat capacity curve, indicating the pairing phase transition from the superfluid phase to the normal phase. This transition is responsible for the S-shape of the heat capacity curve. Furthermore, around T $\approx$ 0.5 MeV, the contribution of $s$ = 4 states begins to increase.



\begin{figure}[ht!]
	\centering
	\includegraphics[scale=.14]{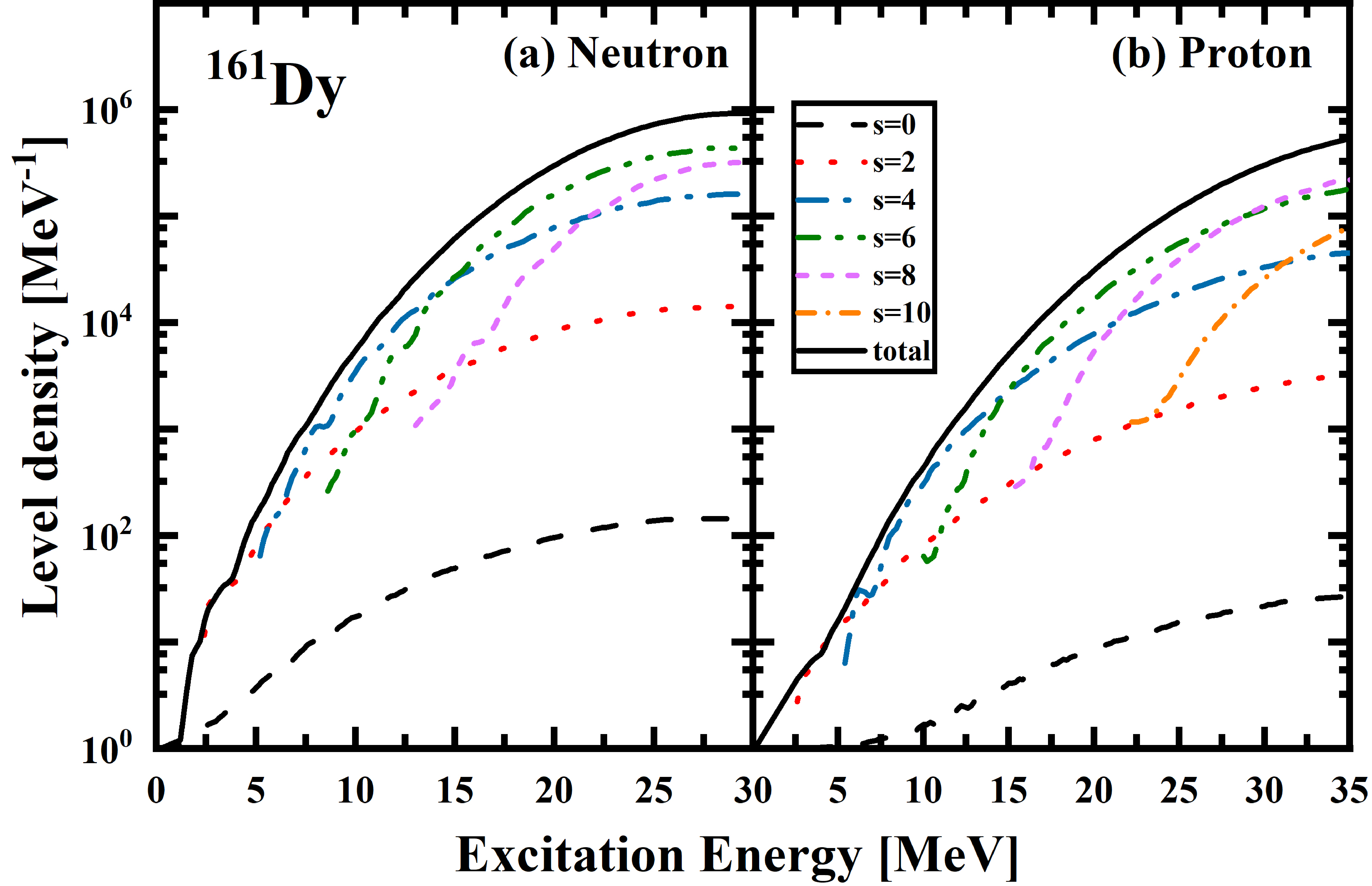}
	\caption{(Color online) Level densities with different seniority numbers $s$ = 0,2,4,6,8 of neutrons (a) and $s$ = 0,2,4,6,8,10 of protons (b) for $^{161} \mathrm{Dy} $ as functions of the excitation energies.}

	\label{fig:example4}
\end{figure}

The level density serves as a valuable tool for characterizing pairing transitions in hot nuclei. To further elucidate the microscopic mechanisms underlying the S-shaped heat capacity curve, we investigate the number of excited states of nuclei in terms of different seniority contributions. Figure~\ref{fig:example4} illustrates the level densities with different seniority numbers $s$ = 0, 2, 4, 6, 8 for neutrons (a) and $s$ = 0, 2, 4, 6, 8, 10 for protons (b) in $^{161} \mathrm{Dy}$ as functions of excitation energies.
From the curves, distinct protuberances are evident for $s$ = 2, 4, 6, and 8 for neutrons, and $s$ = 2 and 4 for protons. Notably, these protuberances are absent at higher seniority numbers. This observation suggests that the initial fluctuations around 4 $\mathrm{MeV}$ are primarily attributed to the contribution of one-pair-broken states. The number of excited states increases rapidly due to the emergence of one-pair-broken states with increasing excitation energies. However, when $T \geq$ 4 $\mathrm{MeV}$, the growth rate of the curves diminishes because no additional one-pair-broken states can be formed. This phenomenon corresponds to the first S-shaped feature in the heat capacity curve.
Subsequently, with the emergence of the second fluctuation, the appearance of two-pair-broken states leads to rapid increase, followed by a decline after 8 $\mathrm{MeV}$ excitation energy for neutrons and 7 $\mathrm{MeV}$ for protons. Other protuberances may indicate additional S-shaped features in the heat capacity curve at higher energies.



\begin{figure}[ht!]
\centering
 \includegraphics[scale=.14]{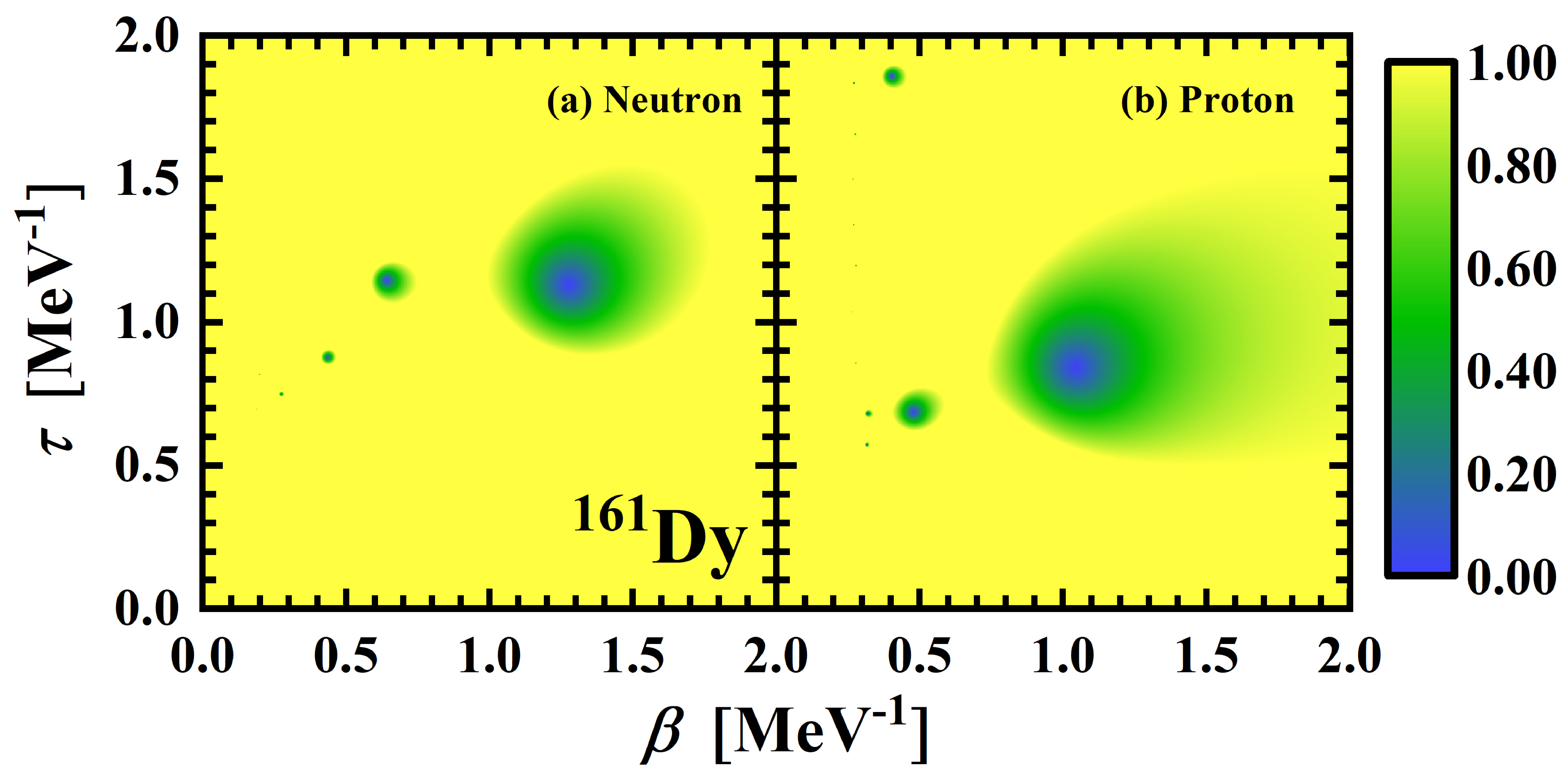}
  \caption{(Color online) The contour plots of the partition function of neutrons and protons for $^{161} \mathrm{Dy} $ in the complex temperature plane.} 
\label{fig:example5}
\end{figure}


\begin{figure}[ht!]
\centering
 \includegraphics[scale=.25]{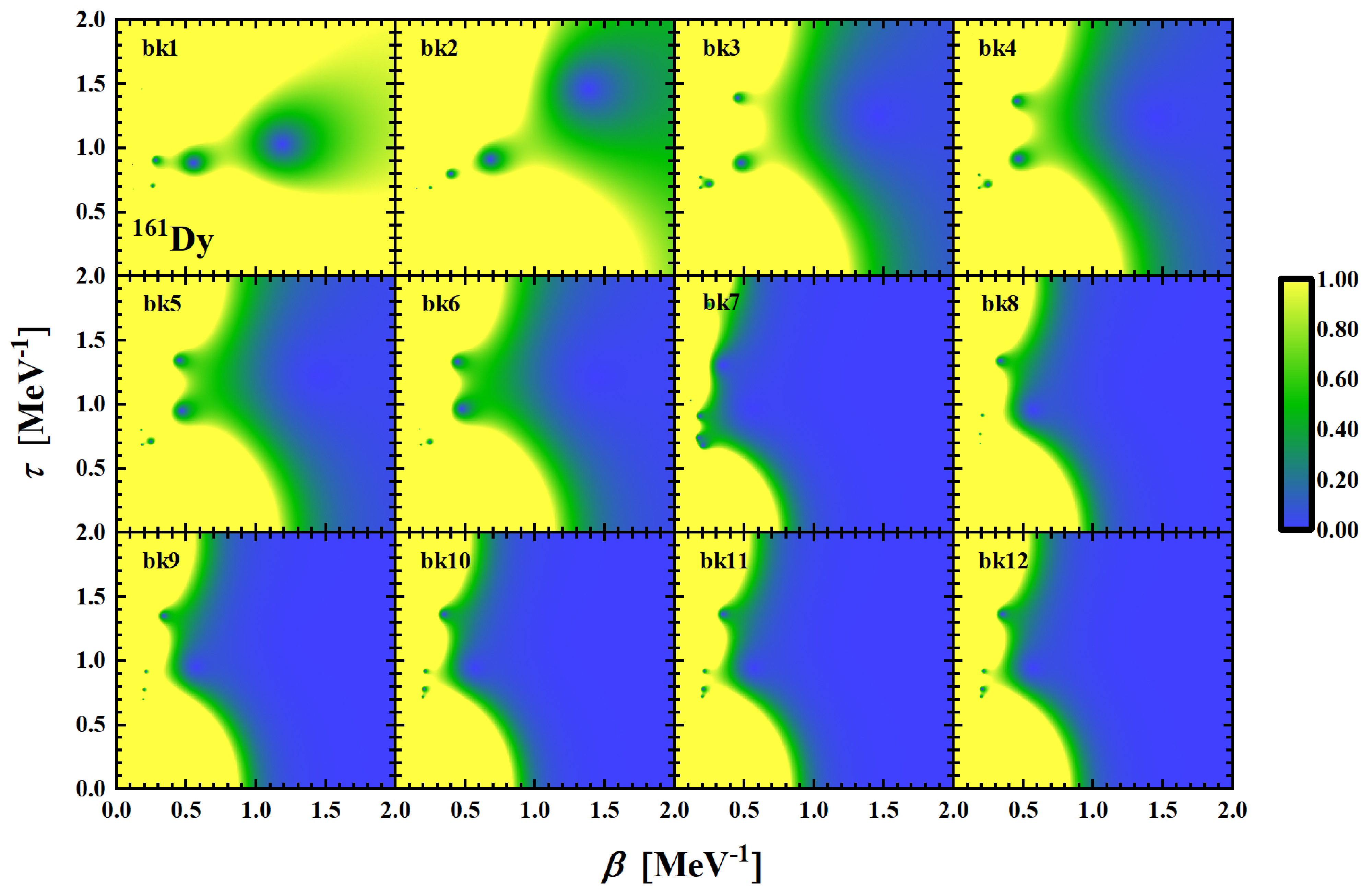}
    \caption{(Color online) The contour plots of the partition function of neutrons for $^{161} \mathrm{Dy} $ with different blocking in the complex temperature plane.} 
\label{fig:example6}
\end{figure}

Figure~\ref{fig:example5} illustrates contour plots of the partition function of neutrons and protons in the complex temperature plane. Zero points at $\beta \approx 1.05 ~ \mathrm{MeV^{-1}}$ for protons and 1.3 $\mathrm{MeV^{-1}}$ for neutrons effectively reflect the critical temperature of the phase transition. The corresponding critical temperature of protons and neutrons are 0.95 MeV and 0.77 MeV, and the experimental value for $^{161} \mathrm{Dy}$ is 0.52 MeV~\cite{PhysRevC.63.021306}. The contour plots of the partition function of neutrons for $^{161} \mathrm{Dy}$ with different blocking scenarios in the complex temperature plane are presented in Fig.~\ref{fig:example6}. It is evident that the partition functions of ``bk1" and ``bk2" contribute the most to the overall partition function. Conversely, patterns of blocking higher energy levels are similar, contributing minimally to the overall partition function. This observation aligns with earlier discussions regarding the effects of different blocking scenarios on heat capacity.
Furthermore, in cases of blocking higher energy levels, the partition function predominantly occupies the zero-point domain. This suggests weakening of the nuclear superfluidity, with the nucleus primarily existing in the normal phase.



\begin{figure}[ht!]
\centering
 \includegraphics[scale=.14]{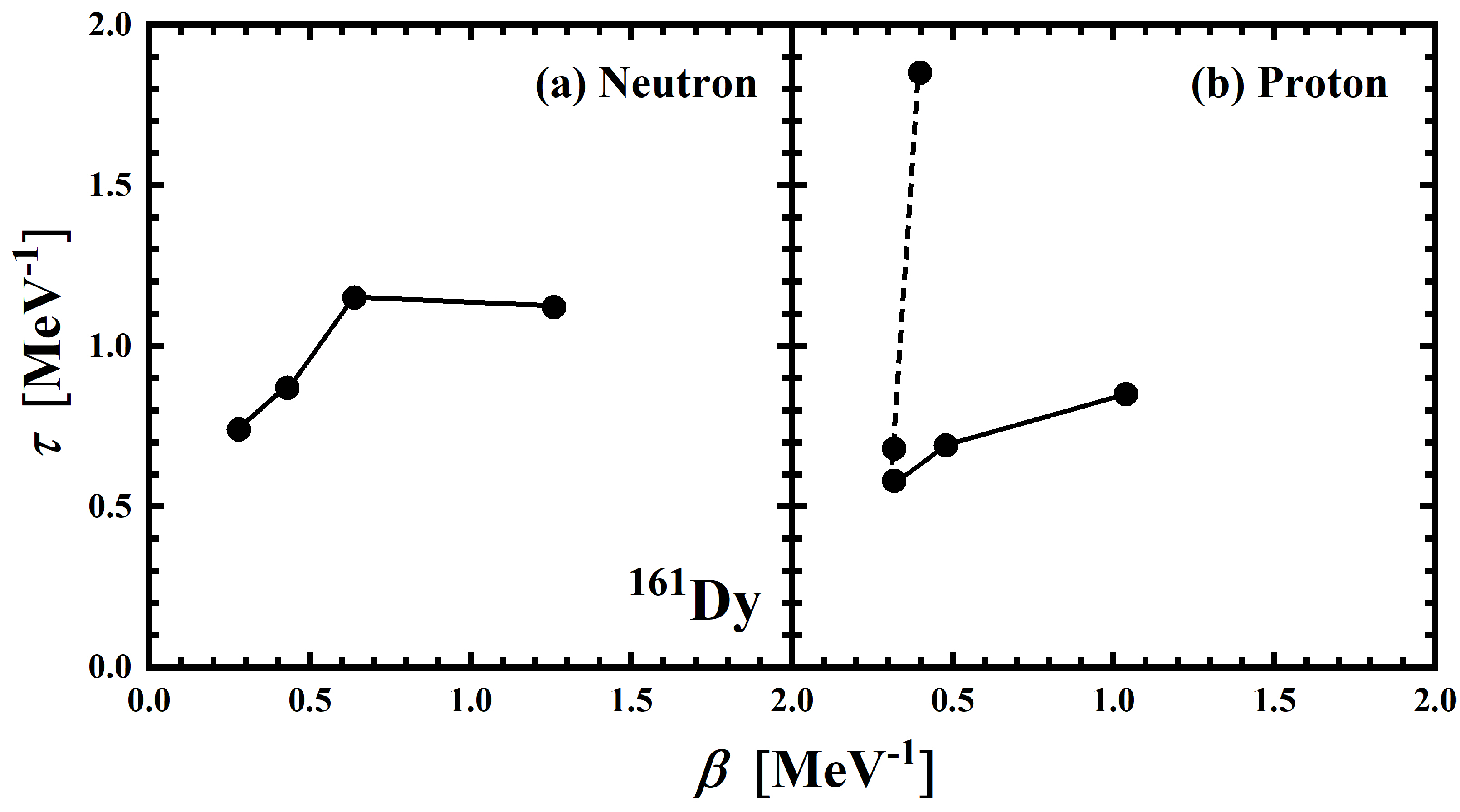}
    \caption{The DOZ of the partition function of the neutrons (a) , protons (b) for $^{161} \mathrm{Dy} $ in the complex temperature plane. The solid line and dashed line represent two kinds of phase transitions.} 
\label{fig:example7}
\end{figure}

According to Fig.~\ref{fig:example5}, the minimums are determined iteratively as the zero points of the partition function. The distributions of zero points of the partition function for neutrons and protons in the complex temperature plane are displayed in Fig.~\ref{fig:example7}, with the zero points of the pairing phase transition indicated by the solid line.
By calculation using Eq.~\ref{eq:12}, the $\alpha$ values corresponding to the neutrons and protons of $^{161} \mathrm{Dy}$ are -2.04 and -5.66, respectively. Based on the phase transition classification method, this indicates a first-order phase transition.

\begin{figure}[ht!]
	\centering
	\includegraphics[scale=.15]{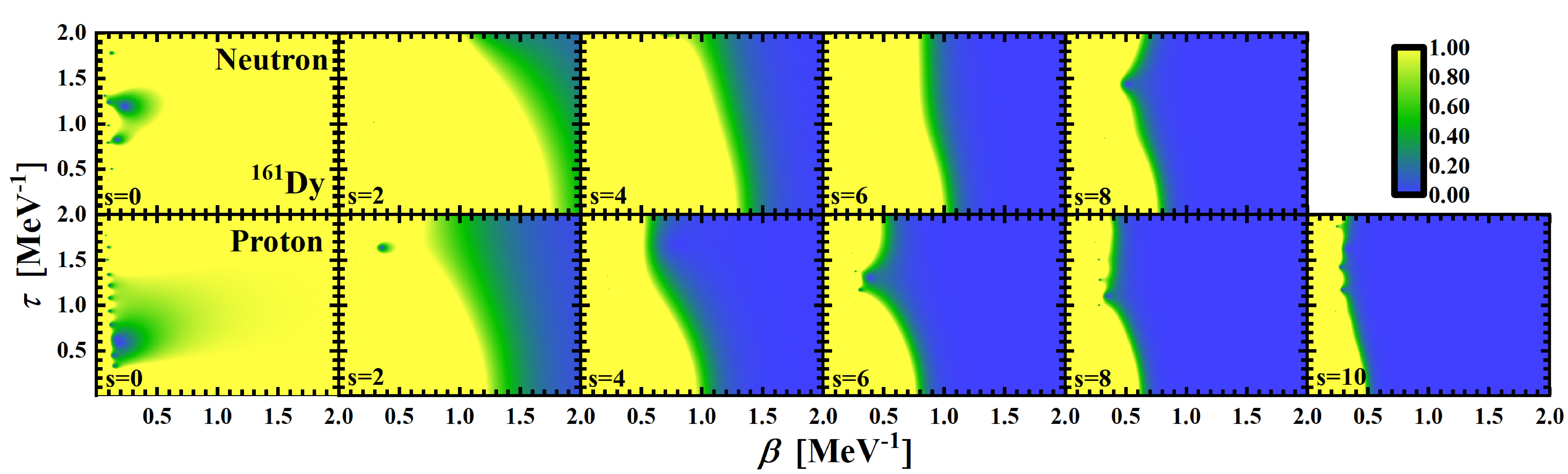}
	\caption{(Color online) The contour plots of the partition function with different seniority numbers s = 0,2,4,6,8,10 of neutrons (top row) and protons (bottom row) for $^{161} \mathrm{Dy} $ in the complex temperature plane.} 
	\label{fig:example8}
\end{figure}

Figure~\ref{fig:example8} illustrates the contour plots of the partition function for $^{161}\mathrm{Dy}$ in the complex temperature plane with different seniority numbers $s$ = 0, 2, 4, 6, 8, and 10, where the upper and lower lines correspond to neutrons and protons, respectively.
In the first column of the seniority number $s$ = 0 (fully paired states), zero points exist only in the region where $\beta$ is very small. Additionally, there are no zero points in the region of 0 to 2 $\mathrm{MeV}$ ($\beta$ $ > $ 0.5 $\mathrm{MeV^{-1}}$), indicating that the nucleus is entirely in the superfluid phase.
As we move to the $s$ = 2 (one pair broken) state, the zero point domain begins to appear with a clear boundary, signifying the coexistence of the superfluid and normal phases. Subsequently, differences emerge between neutrons in odd systems and protons in even systems. The zero domain for neutrons is smaller than that for protons in the $s$ = 2 state, primarily due to the blocking effect of the odd nucleon hindering the evolution of the pairing phase transition.
With increasing seniority number, the zero point domain gradually expands, and the normal phase gradually dominates. The evolution of the pairing phase transition can still be effectively demonstrated. By the time we reach the seniority number $s$ = 8 or 10 (completely broken) state, the image is nearly entirely covered by the zero point domain, indicating that the nucleus is entirely in the normal phase.

\subsection{The Toy Model}

The study above found that the structure of single particle levels affects the heat capacity curve. Under the SLAP framework, single particle levels are derived by Eq.~\ref{eq:1}. In fact, it may be affected by many factors, such as nuclear deformation, shell structure, etc. Since this paper only focuses on the influence of pairing correlation, it is necessary to remove the influence of other factors, so an ideal model can be established. In this study, the ideal models with odd particles and even particles are constructed with controllable pairing strength, and thermodynamic quantities are investigated under the same framework. 
For the even system, a configuration space with 10 levels and 10 particles is selected, with 5 levels above the Fermi surface. The energy levels range from 1 to 10 MeV, and the pairing strength varies from 0 to 2 MeV in increments of 0.1 MeV. The cut-off energy is set to 45 MeV, resulting in a configuration space dimension of $10^4$.
In the case of the odd system, a configuration space with 10 levels and 11 particles is chosen, with 4 levels above the Fermi surface. The single particle levels, pairing strength, and cut-off energy are consistent with the even system. The configuration space dimension is $1.4\times10^4$.\
By employing these ideal models, the study aims to investigate the thermodynamic quantities under the same framework, thus allowing for clearer understanding of the sole influence of pairing correlation on the system.


\begin{figure}[ht!]
	\centering
	\includegraphics[scale=.14]{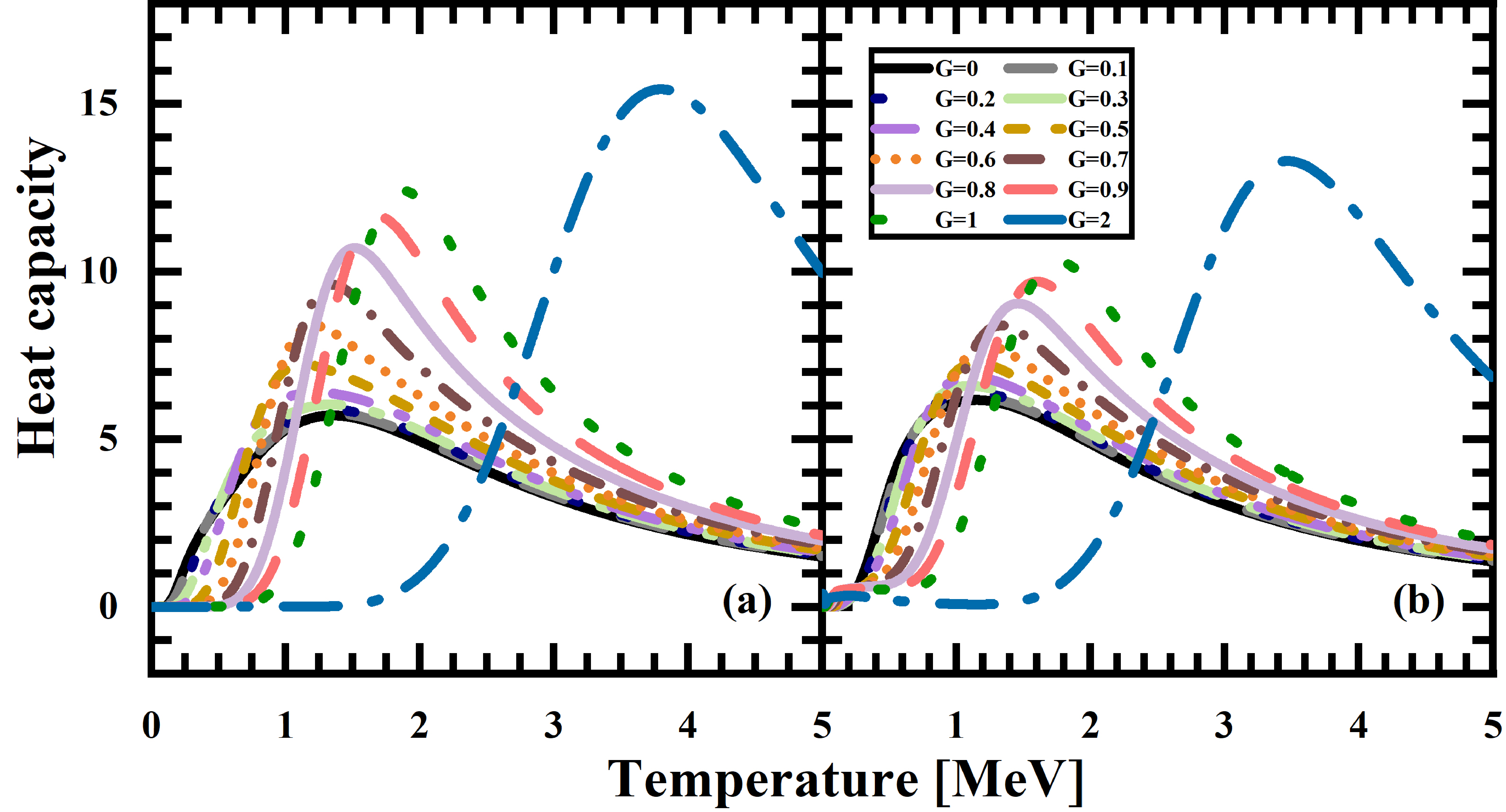}
	\caption{(Color online) The heat capacity with different pairing strengths for the ideal even system (a) and the ideal odd system (b) as functions of temperature.} 
	\label{fig:example9}
\end{figure}

Figure~\ref{fig:example9} illustrates the heat capacity with varying pairing strengths for both the ideal even system (a) and the ideal odd system (b) as functions of temperature. As the pairing strength $G$ increases, the temperatures corresponding to the first and second turning points of the curves, as well as the peak values of the curves, also increase.
In Fig.~\ref{fig:example9} (b), fluctuations are observed in the low-temperature region, similar to the situation discussed for $^{161}\mathrm{Dy}$. These fluctuations are primarily attributed to the pairing re-entrance phenomenon in the odd-nucleon system.



\begin{figure}[ht!]
	\centering
	\includegraphics[scale=.20]{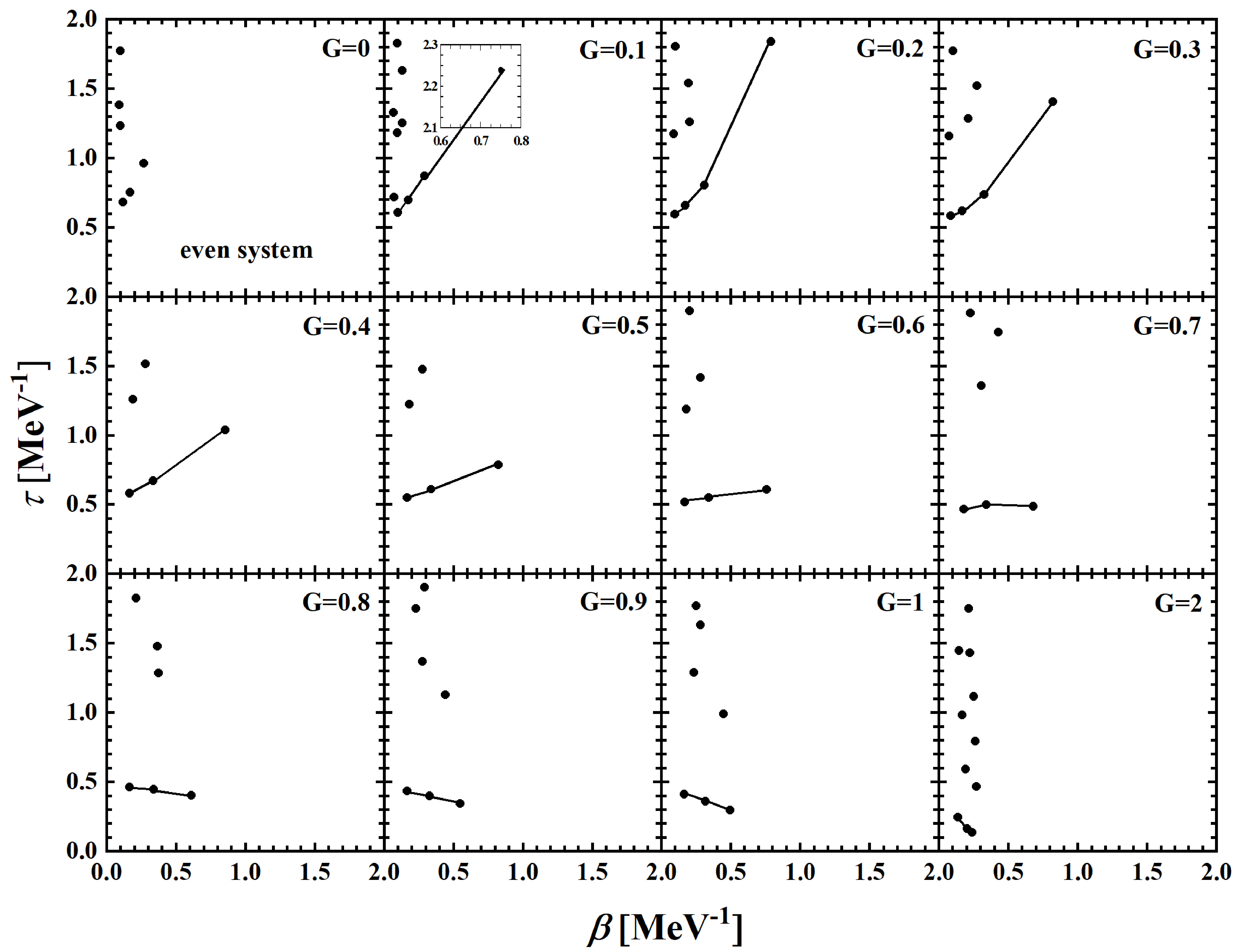}
	\caption{(Color online) The DOZ of the partition function with different pairing strengths for the ideal even system in the complex temperature plane. } 
	\label{fig:example10}
\end{figure}

According to the contour plots of the partition function with varying pairing strengths for the ideal even system in the complex temperature plane, the Distribution of Zeroes (DOZ) of the partition function is displayed in Fig.~\ref{fig:example10}.
When the pairing strength is 0.1, a zero point is observed outside the plotted range, as indicated in the insert figure. In contrast, when the pairing strength is 0 (indicating no pairing correlation), there is no zero point in the corresponding $\beta$ region. However, the heat capacity curve for $G$ = 0 still exhibits an S shape, suggesting that the appearance of an S-shaped heat capacity curve does not solely correspond to the pairing phase transition.
As the pairing strength $G$ increases, the branch of zero points near the real axis undergoes significant changes. The zero point corresponding to the peak value of the heat capacity curves gradually shifts to the region of smaller $\beta$ due to the increasing temperature.

The corresponding $\alpha$ values are calculated and presented in Table \ref{tab:1}. For pairing strengths $G$ $\leq$ 0.7, a first-order phase transition is observed. When $G$ is between 0.8 and 0.9, a higher-order phase transition is identified. Finally, for $G$ = 1, a second-order phase transition is noted. This analysis reveals a nonlinear relationship between the pairing strength and the order of the pairing phase transition, consistent with findings in Ref.~\cite{PhysRevC.76.024319}.


\begin{table}[pt!]
	\renewcommand\arraystretch{1.5}
	\centering
	\caption{The value of $\alpha$ with different pairing strengths for the ideal even system.}
	\label{tab:1} 
	\tabcolsep 8pt 
	
	\begin{tabular}{@{}ccccccccccc@{}}
		\toprule
		$G$ ($\mathrm{MeV}$) & 0.1 & 0.2 & 0.3 & 0.4 & 0.5 & 0.6 & 0.7 & 0.8 & 0.9 & 1.0 \\
		\hline
		$\alpha$ & -3.98 & -4.73 & -5.45 & -3.86 & -5.46 & -10.49 & -39.3 & 7.1 & 2.69 & 0.94 \\ 
		\toprule
	\end{tabular}
	
\end{table}


\begin{figure}[ht!]
	\centering
	\includegraphics[scale=.20]{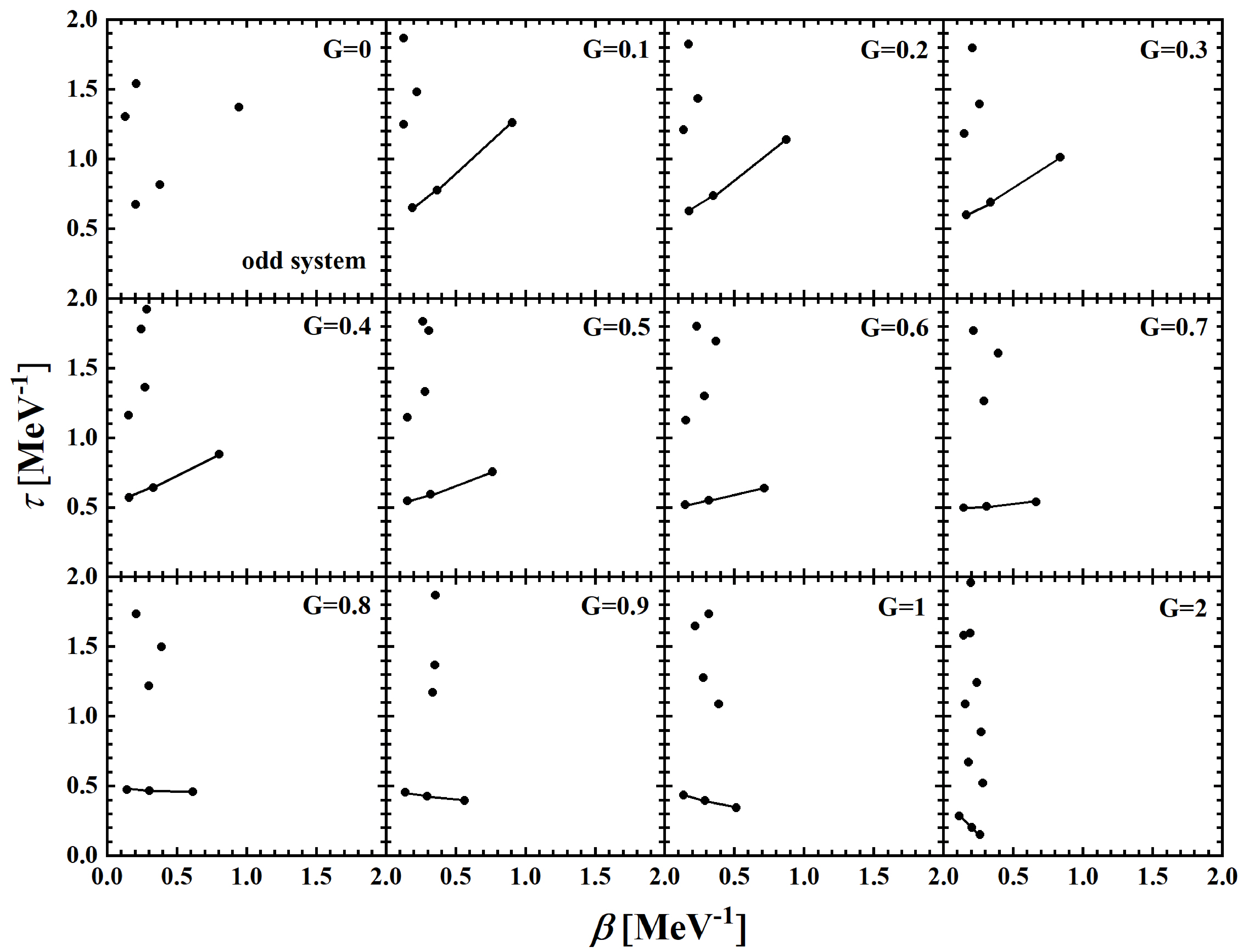}
	\caption{(Color online) The DOZ of the partition function with different pairing strengths for the ideal odd system in the complex temperature plane. } 
	\label{fig:example11}
\end{figure}

Figure~\ref{fig:example11} illustrates the Distribution of Zeroes (DOZ) of the partition function with varying pairing strength for the ideal odd system in the complex temperature plane. Notably, a branch is observed when the pairing strength is 0. Presence of this branch is primarily attributed to changes in the zero point distribution caused by the blocking effect, and it does not correspond to a real pairing phase transition. This behavior contrasts with the case of even systems.

\begin{table}[pt]
	\renewcommand\arraystretch{1.5}
	\centering
	\caption{The value of $\alpha$ with different pairing strengths for the ideal odd system.}
	\label{tab:2} 
	\tabcolsep 8pt 
	
	\begin{tabular}{@{}ccccccccccc@{}}
		\toprule
		$G$ ($\mathrm{MeV}$) & 0.1 & 0.2 & 0.3 & 0.4 & 0.5 & 0.6 & 0.7 & 0.8 & 0.9 & 1.0 \\
		\hline
		$\alpha$ & -3.4 & -3.66 & -3.98 & -4.66 & -4.61 & -8.37 & -18.59 & 35.01 & 7.2 & 3.2 \\ 
		\toprule
	\end{tabular}
	
\end{table}

Based on the zero distribution depicted in Fig.~\ref{fig:example11}, the corresponding $\alpha$ values are calculated and presented in Table~\ref{tab:2}. For pairing strengths $G$ $\leq$ 0.7, a first-order phase transition is identified. However, for $G$ between 0.8 and 1, a higher-order phase transition is observed. This analysis demonstrates that the pairing strength also influences the pairing phase transition in the odd system.

\section{Summary}
\label{sec:4}

In summary, this work investigates the pairing phase transition in the odd-A nucleus $^{161}\mathrm{Dy}$ using the covariant density functional theory and the shell-model-like approach. Thermodynamic quantities are evaluated within the canonical ensemble theory, and the phase transition order is determined using a classification scheme based on the distribution of zeros of the partition function in the complex temperature plane. The impact of different pairing strengths on the pairing phase transition is also examined.

The heat capacity curve of $^{161} \mathrm{Dy}$ exhibits an S shape, indicating a pairing phase transition from the superfluid phase to the normal phase. Extending the calculation to the complex temperature plane reveals that the pairing phase transition is of first order. Blocking the Fermi surface and the first level above it significantly contribute to the total partition function, highlighting the importance of considering the blocking of the nearest levels above the Fermi surface in the study of odd-A nuclei. The contour plots of the partition function with different seniority show that the blocking effect of the odd nucleon hinders the evolution of the pairing phase transition.

Furthermore, ideal models with odd and even particles are constructed to explore the influence of different pairing strengths on the pairing phase transition in the complex temperature plane. In the ideal even system, the absence of pairing correlation does not lead to a zero point corresponding to the critical temperature in the large $\beta$ region, yet the heat capacity curve still exhibits an S shape. This suggests that the appearance of an S-shaped heat capacity curve does not exclusively indicate a pairing phase transition. Conversely, in the ideal odd system, the presence of a zero point branch in the partition function plot is due to the blocking effect. The analysis also demonstrates that the pairing strength has nonlinear influence on the pairing phase transition. In the ideal even system, the phase transition order changes from first order to higher order and then to second order as the pairing strength increases. For the ideal odd system, the phase transition order changes from first order to higher order with increasing pairing strength. In the near future research, we will focus more on the experimental spectroscopic properties and other properties to study the pairing.

\begin{acknowledgments}
This work was supported by National Natural Science Foundation of China under Grant No. 11775099 and the Jiangnan University Basic Research Program (JUSRP202406002).
\end{acknowledgments}

\bibliographystyle{apsrev4-1}
\bibliography{heat}

\end{document}